\documentclass[a4paper,12pt]{article}
\usepackage{epsfig}

\begin{document}

\begin{center}
{\Large{\bf Evaluation of the $\pi\pi$ scattering amplitude in the
$\sigma$-channel at finite density}} 
\end{center}

\vspace{0.4cm}

\begin{center}
{\large{D. Cabrera, E. Oset and M.J. Vicente Vacas}}
\end{center}

\begin{center}
{\small{\it Departamento de F\'{\i}sica Te\'orica and IFIC, \\
Centro Mixto Universidad de Valencia-CSIC, \\
Ap. Correos 22085, E-46071 Valencia, Spain}}
\end{center}

\vspace{0.8cm}

\begin{abstract}
The $\pi\pi$ scattering amplitude in the
$\sigma$-channel is studied at finite baryonic density in the
framework of  a chiral unitary approach which  successfully reproduces the
meson meson phase shifts and  generates the $f_0$ and $\sigma$ resonances in
vacuum. We address here a new variety of mechanisms recently suggested to
modify the $\pi\pi$ interaction in the medium, as well as the role of the
$s-$wave selfenergy, in addition to the $p-$wave, in the dressing of the pion
propagators.
\end{abstract}

\section{Introduction}

In the last years, several theoretical approaches have predicted strong medium
effects on the pion pion interaction in the scalar isoscalar ($\sigma$) channel.
 
In Ref.  \cite{Hatsuda:1999kd}, Hatsuda et al. studied the $\sigma$ propagator
in the linear $\sigma$ model and found an enhanced and narrow spectral 
function  near the $2\pi$ threshold  caused by the partial restoration of the
chiral symmetry, where $m_\sigma$ would approach $m_\pi$. The same
conclusions were reached using the nonlinear chiral Lagrangians  in Ref.
\cite{Jido:2000bw}.

Similar results, with large enhancements in the $\pi\pi$ amplitude around the 
$2\pi$ threshold, have been found in a quite different approach by studying the
$s-$wave, $I=0$ $\pi\pi$ correlations in nuclear matter
\cite{Schuck:1988jn,Rapp:1996ir,Aouissat:1995sx,Chiang:1998di}. In these cases
the modifications of the $\sigma$ channel are induced by the strong $p-$wave
coupling of the pions to the particle-hole ($ph$) and $\Delta$-hole ($\Delta
h$) nuclear excitations.  It was pointed out in 
\cite{Aouissat:2000ss,Davesne:2000qj} that this attractive $\sigma$ selfenergy
induced by the $\pi$ renormalization in the nuclear medium could be
complementary to additional $s$-wave renormalizations of the kind discussed in 
\cite{Hatsuda:1999kd,Jido:2000bw} calling for even larger effects.

On the experimental side, there are also several results showing strong medium
effects in the $\sigma$ channel at low invariant masses in the $A(\pi,2\pi)$ 
\cite{Bonutti:1996ij,Bonutti:1998zw,Camerini:1993ac,bonutti,Starostin:2000cb}
and $A(\gamma,2\pi)$  \cite{Messchendorp:2002au} reactions. At the moment, the
cleanest  signal probably corresponds to the $A(\gamma,2\pi^0)$ reaction, which
shows large density effects that had been predicted in both shape and size in
Ref. \cite{Roca:2002vd}, using a model for the $\pi\pi$ final state interaction
along the lines of the present work. Note, however, that a part of the spectrum
modification could be due to quasielastic collisions of the pion
\cite{Muhlich:2004zj}.

Our aim in this paper is to study the  $\pi\pi$ scattering in the scalar
isoscalar ($\sigma$) channel at finite densities in the context of the model
developed in 
\cite{Dobado:1990qm,Dobado:1993ha,Oller:1998ng,Oller:1999hw,Oller:1999zr,Oller:1997ti}.
These works, which provide an economical and successful description of a
wide range of hadronic phenomenology, use as input the lowest orders of the
Lagrangian of Chiral Perturbation Theory ($\chi PT$)  \cite{Gasser:1985ux} and calculate
meson meson scattering  in a coupled channels unitary way. 
Some nuclear medium effects, namely the $p-$wave coupling of the pions to the
particle hole ($ph$) and Delta hole ($\Delta h$) excitations, were
implemented in this framework in Refs.  \cite{Chiang:1998di,Oset:2000ev}. As in
other approaches, large medium effects were found as reflected in the imaginary
part of the  $\pi \pi$ scattering amplitude which showed a clear shift of
strength towards low  energies as the density increases.
Although this model was able to predict the size of the medium effects on the
$(\gamma, 2\pi)$ reaction \cite{Roca:2002vd}, it was pointed out that some
probably large contributions related to nucleon tadpole diagrams
\cite{Jido:2000bw} and some vertex corrections \cite{Meissner:2001gz} were
missing. In this work, we will include those pieces and analize its influence in
the $\pi\pi$ scattering amplitude at finite nuclear densities.

In the next section we present, for the sake of completeness, a brief 
description of the model used for the $\pi \pi$ interaction both in vacuum and
in a dense medium, which is already published elsewhere \cite{Chiang:1998di,Oset:2000ev}. 
In Section 3 we consider further contributions to the $\pi\pi$ interaction in
the nuclear medium, associated to higher order terms in the chiral Lagrangian
than those included in Refs. \cite{Chiang:1998di,Oset:2000ev}, and some baryonic
vertex corrections advocated in Ref. \cite{Meissner:2001gz}.

\section{$\pi \pi$ interaction}

In this section we summarize the method of Ref.  \cite{Oller:1997ti}
for $\pi \pi$ interaction in vacuum and Refs.  \cite{Chiang:1998di,Oset:2000ev}
for the nuclear medium effects. Additional information on this and related 
approaches for different spin isospin channels can be found in Refs. 
\cite{Oller:1997ti,Oller:1998ng,Oller:1999hw,Nieves:2000bx}.

\subsection{Vacuum}

 The basic idea is to solve a  Bethe Salpeter (BS) equation, which guarantees
unitarity, matching the low energy results to $\chi PT$ predictions. We
consider two coupled channels, $\pi \pi$ and $K \bar{K}$ and neglect the $\eta
\eta $ channel which is not relevant at the low energies  we are interested
in. 

\begin{figure}
\begin{center}
 \epsfig {figure=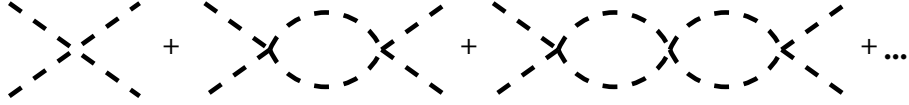,width=12.cm}
 \caption{Diagrammatic representation of the Bethe Salpeter equation.}
 \label{fig:BSF}
\end{center}
\end{figure}

    The BS equation is given by 
\begin{equation}
\label{eq:BS}
T=V+VGT.
\end{equation}   
Eq. (\ref{eq:BS})  is a matrix integral equation  which 
involves the two mesons one loop divergent integral (see Fig.~\ref{fig:BSF}),  
where  $V$ and $T$ appear off shell. However, for this channel both functions 
can be factorized on shell out of the integral. The remaining off shell part 
can be absorbed by a renormalization of  the coupling constants
as it was shown in Refs. \cite{Oller:1997ti,Nieves:1999hp}. 
 Thus, the BS equation becomes purely algebraic and the 
$VGT$ term, originally inside the loop integral, becomes then the product 
of $V$, $G$ and $T$, with $V$ and $T$ the on shell amplitudes independent
of the integration variables, and  $G$  given by  the  expression  
\begin{equation}
G_{ii}(P) = i \int \frac{d^4 q}{(2 \pi)^4}
\frac{1}{q^2 - m_{1i}^2 + i \epsilon} \; \; 
\frac{1}{(P - q)^2 - m_{2i}^2 + i \epsilon}
\end{equation}
where $P$ is the momentum of the meson meson system. This 
integral is regularized with a cut-off ($\Lambda$) adjusted to optimize the fit
to the $\pi\pi$ phase shifts ($\Lambda=1.03$ GeV).
The potential $V$ appearing in the BS equation is taken from 
the lowest order chiral Lagrangian
\begin{equation}
{\cal L}_2 = \frac{1}{12 f^2} \langle (\partial_\mu \Phi \Phi - \Phi \partial_\mu
\Phi)^2 + M  \Phi^4 \, \rangle
\end{equation}

\noindent
where the symbol $\langle \rangle$ indicates the trace in flavour space,
$f$ is the pion decay constant and $\Phi$, $M$ are the pseudoscalar meson 
and mass $SU(3)$ matrices.
This model reproduces well phase shifts and inelasticities up to about 
1.2 GeV. The $\sigma$ and $f_0 (980)$ resonances appear as poles of the 
scattering amplitude in $L=0$, $I=0$. 
The coupling of channels is essential to produce the 
$f_0 (980)$ resonance, while the $\sigma$ pole is little
affected by the coupling of the pions to $K \bar{K}$
 \cite{Oller:1997ti}. 

\subsection{\label{sec:nucmed}The nuclear medium}

As we are mainly interested in the low energy region, which is not very
sensitive to the kaon channels, we will only consider the nuclear medium 
effects on the pions. The main changes of the pion propagation in the 
nuclear medium come from the $p-$wave selfenergy, produced basically 
by the coupling of pions to particle-hole ($ph$) and Delta-hole 
($\Delta h$) excitations. For a pion of momentum $q$ it is given by 
\begin{equation}
\label{eq:self}
  \Pi(q)= {{\left({D+F}\over{2f}\right)^2 \vec q\,^2 U(q)}
            \over
           {1-\left({D+F}\over{2f}\right)^2 g' U(q)}}
\end{equation}
with $g'=0.7$ the Landau-Migdal parameter, $U(q)$ the  Lindhard function and
$(D+F)=1.257$. The expressions for the Lindhard functions  are taken
from Ref. \cite{Oset:1990ey}. 

\begin{figure}[htb]
 \begin{center}
\epsfig{height=2.2cm,width=12.1cm,angle=0, figure=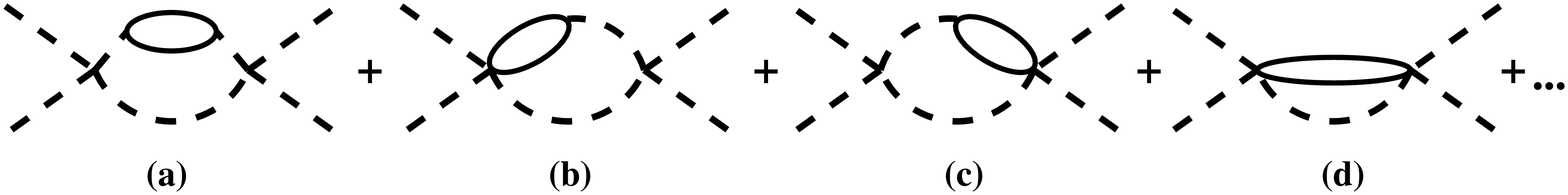}
\caption{Terms of the meson meson scattering amplitude accounting for
$ph$ and $\Delta h$ excitation.}
 \label{fig:BSF2}
 \end{center}
\end{figure}

Thus, the in-medium BS equation  will include the 
diagrams of Fig. \ref{fig:BSF2} where the solid line bubbles represent
the $ph$ and $\Delta h$ excitations.
In fact, as it was shown in  \cite{Chanfray:1999nn}, the contact terms with
the $ph$ ($\Delta h$) excitations of diagrams (b-d) cancel the 
off-shell contribution  from the meson meson vertices in the term of Fig.
\ref{fig:BSF2}(a). 
Hence, we just need to calculate the diagrams of the free type 
(Fig. \ref{fig:BSF}) and those of Fig. \ref{fig:BSF2}(a) with the amplitudes
factorized on shell. Therefore, at first order in the baryon density, we are left 
with simple meson propagator corrections which can be readily incorporated by 
changing the meson vacuum propagators by the in medium ones.

\begin{figure}[htb]
 \begin{center}
\epsfig{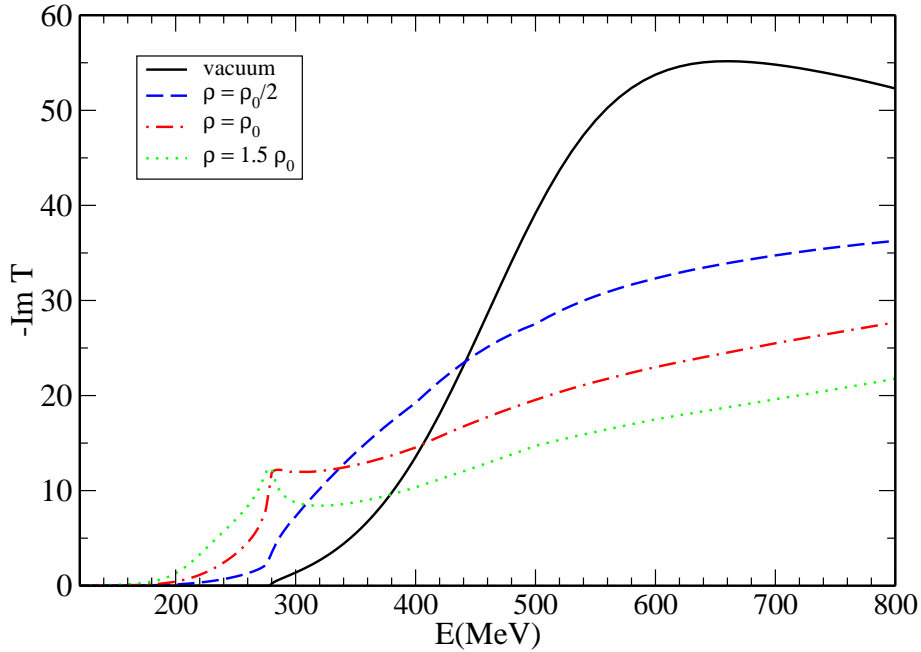}
 \caption{Imaginary part of the $\pi\pi$ scattering amplitude at several
 densities.}
 \label{fig:MED1}
\end{center}
\end{figure}

 The $\pi\pi$ scattering amplitude  obtained using this model
exhibits a strong shift towards low energies. In Fig. \ref{fig:MED1}, we show
the imaginary part of this amplitude for several densities. Quite similar
results have been found using different models 
 \cite{Aouissat:1995sx} and it has been suggested
that this accumulation of strength, close to the pion threshold, could reflect
a shift of the $\sigma$ pole which would approach the mass of the pion. 
Other pion selfenergy contributions related to $2ph$ excitations, and thus 
proportional to $\rho^2$, can be incorporated in the pion propagator. As we are
most interested in the region of low energies we can take as estimation the
corresponding piece of the optical potentials obtained from pionic atoms data,
following the procedure of Ref. \cite{Chiang:1998di} and substituting in Eq.
(\ref{eq:self})
\begin{equation}
\left({D+F}\over{2f}\right)^2  U(q)
\;\longrightarrow\; 
\left({D+F}\over{2f}\right)^2  U(q) -4\pi C_0^* \rho ^2
\label{eq:self2}
\end{equation}
with $\rho$ the nuclear density and $C_0^*=(0.105+i 0.096) m_\pi^{-6}$.
Its effects are small except at large densities as can be appreciated by
comparing  Fig. \ref{fig:MED1}, with Fig. 7 of Ref. \cite{Chiang:1998di}
where this piece is included.

\section{Further contributions}

\subsection{\label{sec:tadpoles}Higher order tadpole and related terms}

\begin{figure}[htb]
 \begin{center}
\epsfig{width=6cm,figure=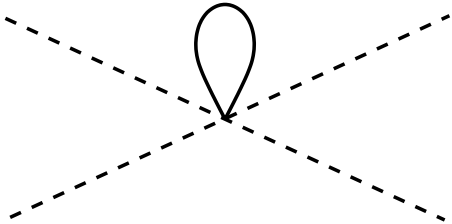}
 \caption{Nucleon tadpole term correction to the $\pi\pi$ interaction.}
 \label{fig:NUCTAD}
 \end{center}
\end{figure}

The chiral Lagrangian generates tadpole terms that could contribute to the
pion selfenergy and also in the form of vertex corrections as in Fig.
\ref{fig:NUCTAD}. At the lowest order these terms vanish in isospin symmetric
nuclear matter \cite{Oset:2000ev}. However, at next order there are terms which
provide some contribution. The complete structure of the higher order
Lagrangian adapted to the $\pi N$ system can be seen in \cite{Bernard:1995dp}.
The medium corrections associated to these new Lagrangian terms in the $\pi$ nucleus
interaction were studied in \cite{Thorsson:1995rj} and interpreted in terms of
changes  of the time and space components of $f$ and changes of the pion mass
in the medium. Further developments in this direction are done in
\cite{Meissner:2001gz}.

The repercussion of these terms in $\pi\pi$ scattering in the nuclear medium
has been considered in \cite{Jido:2000bw} and we follow here the same steps. We
start from the second order $\pi N$ Lagrangian relevant for the isoscalar
sector
\begin{eqnarray}
\label{LpiN2}
{\cal L}_{\pi N}^{(2)} & = &  c_3 \bar{N} (u_{\mu}  u^{\mu}) N
 + (c_2 - {g_A^2 \over 8 m_N})  \bar{N}({\rm v}_{\mu}  u^{\mu})^2 N \nonumber \\
 &  & +  c_1 \bar{N}N {\rm Tr} (U^{\dagger}  \chi + \chi^{\dagger} U)
      + \cdot \cdot \cdot ,
\end{eqnarray}
where $u_{\mu} = i u^{\dagger} \partial_{\mu} U u^{\dagger}$, with $U =
u^2 = {\rm exp}(i \tau^a \phi^a /f)$ in the $SU(2)$ formalism used
there, ${\rm v}_{\mu}$ is the four velocity of the nucleon, $g_A$ the axial
charge of the nucleon and $\chi = {\rm diag}(m_{\pi}^2,m_{\pi}^2)$. 
The pion nucleon amplitude obtained from the
Lagrangian in Eq. (\ref{LpiN2}) is
\begin{eqnarray}
\label{TpiNfromLpiN2}
t_{\pi N} &=& \frac{4 c_1}{f^2} m_{\pi}^2 - \frac{2 c_2}{f^2} (q^0)^2 
- \frac{2 c_3}{f^2} q^2
\nonumber \\
&=& ( \frac{4 c_1}{f^2} m_{\pi}^2 - \frac{2 c_2}{f^2} \omega(q)^2 
- \frac{2 c_3}{f^2} m_{\pi}^2 )
\nonumber \\
& & - \frac{2 c_2 + 2 c_3}{f^2} (q^2-m_{\pi}^2) 
= t_{\pi N}^{on} + t_{\pi N}^{off} \,\,\, ,
\end{eqnarray}
where in the last part of the equation we have separated what we call the
on-shell part and the off-shell part of the amplitude (term with
$(q^2-m_{\pi}^2)$). This $s-$wave $\pi N$ interaction produces a modification of
the pion propagator which we shall consider later in the solution of the Bethe
Salpeter equation in the medium.

\begin{figure}[htb]
 \begin{center}
\epsfig{width=8cm,figure=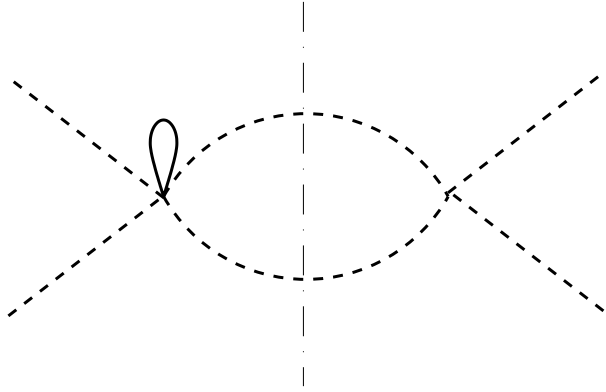}
 \caption{$\pi\pi$ rescattering diagram with tadpole vertex correction showing
 the $\pi\pi$ cut.}
 \label{fig:NUCTADLOOP}
 \end{center}
\end{figure}

In \cite{Jido:2000bw}
and \cite{Thorsson:1995rj} the medium effects are recast at
the mean field level in terms of a medium Lagrangian given by
\begin{eqnarray}
\label{mean-f}
\langle {\cal L} \rangle
 & = & ( {f^2 \over 4} + {c_3 \over 2} \rho)\  {\rm Tr}
 [\partial_{\mu} U \partial^{\mu} U^{\dagger}] \nonumber \\
 &  &  + \ \  ( {c_2 \over 2} - {g_A^2 \over 16 m_N})\ \rho \  {\rm Tr}
 [\partial_0 U \partial_0 U^{\dagger}] \nonumber \\
 &  & +  \ \
  ({ f^2 \over 4} + {c_1 \over 2} \rho)
 \ {\rm Tr} (U^{\dagger}  \chi + \chi^{\dagger} U) \,\,\, .
\end{eqnarray}
The different corrections to the $\pi\pi$ scattering
amplitude coming from the
$\partial_{\mu} U \partial^{\mu} U^{\dagger}$, $\partial_0 U \partial^0
U^{\dagger}$ terms and the mass term in Eq. 
(\ref{mean-f}) ($c_3$, $c_2$ and $c_1$ terms) are given by
\begin{eqnarray}
\label{correc_derivative}
\delta t_{\pi\pi}^{(t)}
&=&- \frac{1}{f^2} \lbrace \frac{2 c_3}{f^2} \rho (s-\frac{4}{3}m_{\pi}^2) 
+ \frac{2 c_2}{f^2} \rho (s-\frac{1}{3} \sum_i \omega_i(q)^2) +
\frac{c_1}{f^2}\rho\frac{5}{6} m_{\pi}^2 \rbrace
\nonumber \\
& &+ \frac{1}{f^2} \lbrace \frac{2 c_3}{f^2} \rho \frac{1}{3} \sum_i
(q_i^2-m_{\pi}^2) + \frac{2 c_2}{f^2} \rho \frac{1}{3} \sum_i
(q_i^2-m_{\pi}^2) \rbrace \,\,\, ,
\end{eqnarray}
where we have also separated the on-shell part from the off-shell part. These
are the corrections coming from the many body tadpole diagram of Fig.
\ref{fig:NUCTAD}, 
which are included in the $\rho$ dependent terms of Eq. (\ref{mean-f}).
Note that in the chiral unitary approach that we follow, the external legs are
placed on shell ($q_i^2=m_{\pi}^2$). This is the case even when the diagrams
appear in loops, as in Fig. \ref{fig:NUCTADLOOP}, since the underlying physics
is the use of a dispersion relation using the $N/D$ method
\cite{Oller:1998zr,Oller:2000fj}
which determines the diagram contribution in terms of its imaginary part. In
the case of Fig. \ref{fig:NUCTADLOOP} the cut corresponds to two free pions on
shell, like in the vacuum. Hence, we shall use only the on shell part of the
correction of Eq. (\ref{correc_derivative}).

\begin{figure}[htb]
 \begin{center}
\epsfig{width=10cm,figure=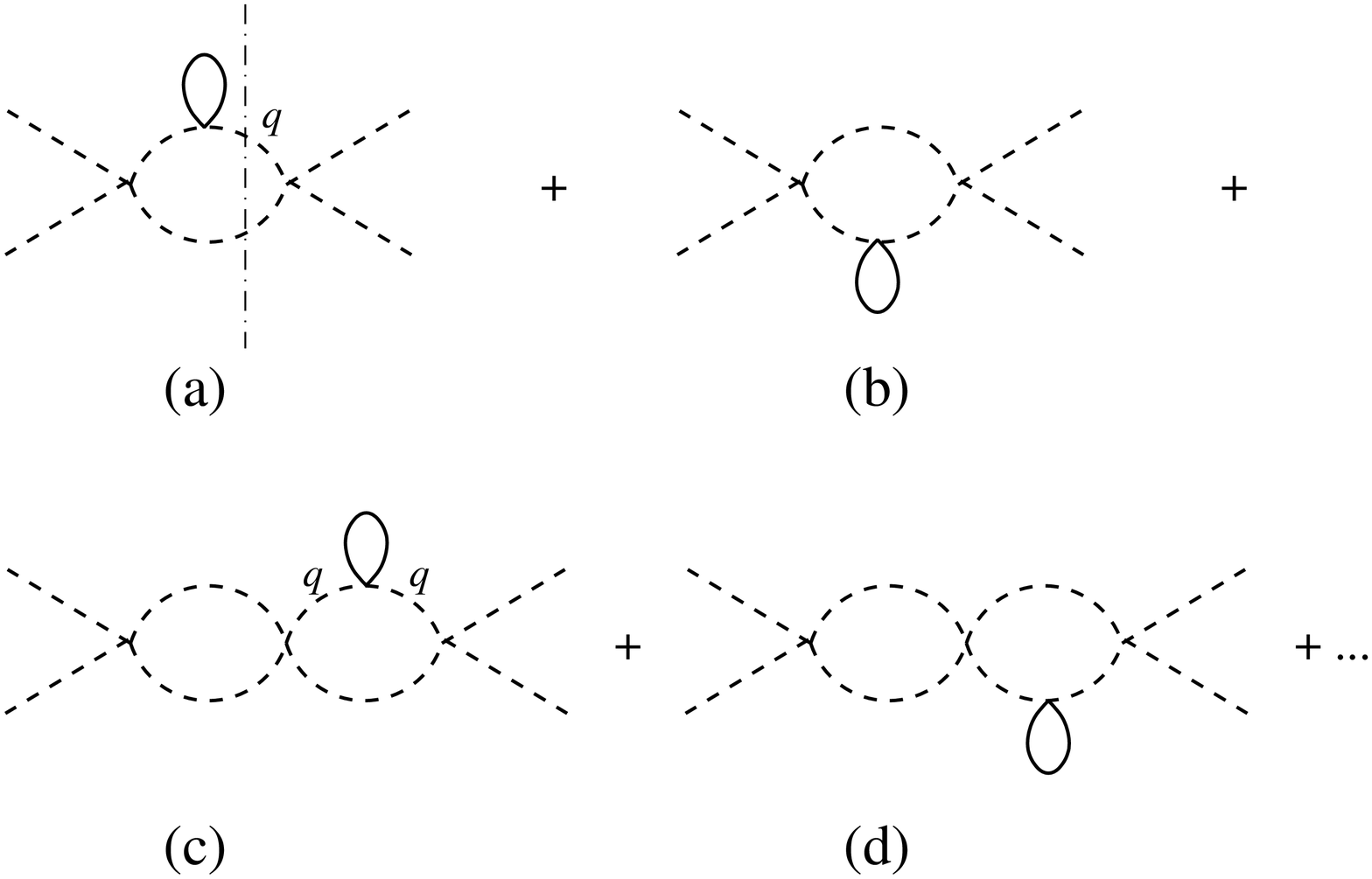}
 \caption{Nucleon tadpole correction in the pion propagator.}
 \label{fig:NUCTADLOOPSERIES}
 \end{center}
\end{figure}

As mentioned before, at the same time, when solving the Bethe-Salpeter equation, 
we have also to take into account the $s-$wave selfenergy
insertion from the Lagrangian of Eq. (\ref{LpiN2}) in the pion propagators as
depicted in Fig. \ref{fig:NUCTADLOOPSERIES}.
This is easily accounted for, at lowest order in $\rho$, adding to each pion
propagator, $D_{\pi}$, the correction $D_{\pi} t_{\pi N} \rho D_{\pi}$. A
technically simple way to account for that is to add to the 
scalar isoscalar $\pi\pi$ vertex from ${\cal L}_2$, $t_{\pi\pi}$, the correction
\begin{equation}
\label{swaveinsertion}
\delta t_{\pi\pi}^{(s)} = 
(t_{\pi N}^{on}+t_{\pi N}^{off}) \rho \frac{1}{q^2-m_{\pi}^2} t_{\pi\pi}
\end{equation}
for the two pion propagator lines to the left of the $\pi\pi$ vertex.
Now the separation of the on-shell and off-shell parts of $t_{\pi N}$ is most
useful since the pion propagator in Eq.
(\ref{swaveinsertion}) is cancelled out by the $(q^2-m_{\pi}^2)$ factor of the
off-shell part of $t_{\pi N}$.
Thus we have
\begin{equation}
\label{swaveinsertion2}
\delta t_{\pi\pi}^{(s)} = t_{\pi N}^{on} \, \rho \, \frac{1}{q^2-m_{\pi}^2}
t_{\pi\pi} - \frac{2 c_2 + 2 c_3}{f^2} \, \rho \, t_{\pi\pi} \,\,\, .
\end{equation}
This means that with the Lagrangian used, on top of the corrections in the loops
from the $s-$wave (on-shell) pion selfenergy, we have an additional correction
(second term of Eq. (\ref{swaveinsertion2})) of the same topology as
the tadpole term considered before.
Considering the pion
selfenergy insertion in either of the two pion propagators, we obtain for this
\begin{equation}
\label{swaveinsertion3}
\delta t_{\pi\pi}^{(st)} = -\frac{4c_2 + 4c_3}{f^2} \, \rho \, t_{\pi\pi} \,\,\, .
\end{equation}
Thus, we are left with the usual contribution in the pion loops of the ordinary
on-shell $s-$wave pion selfenergy, plus the tadpole correction of Eq.
(\ref{correc_derivative}), plus the tadpole equivalent of Eq.
(\ref{swaveinsertion3}).

There are still further contributions belonging to the same family. Indeed, the
$t_{\pi\pi}$ amplitude in the scalar isoscalar channel,
\begin{equation}
\label{tpipiatrhozero}
t_{\pi\pi} = -\frac{1}{f^2} (s-\frac{m_{\pi}^2}{2} -
\frac{1}{3}\sum_i(q_i^2-m_{\pi}^2)) \,\,\,,
\end{equation}
is also split in on- and off-shell parts. In \cite{Chiang:1998di,Chanfray:nn} it
was shown that the off-shell pieces could be removed from the loop calculations
for both the free pion case and the pion with a $p-$wave selfenergy. 
However, the diagrams in Fig. \ref{fig:NUCTADLOOPSERIES}(a) have one free pion and a
pion with a $s-$wave medium selfenergy insertion, hence the imaginary part of the
two-pion loop is not the same 
as in the mentioned cases.
It is again easy to take into
account this correction and we have, from the $s-$wave selfenergy insertions in
the pion propagators
\begin{equation}
\label{swaveinsertion4}
\delta t_{\pi\pi}^{(so)} = t_{\pi N}\rho \frac{1}{q^2-m_{\pi}^2} \frac{1}{3 f^2} (q^2-m_{\pi}^2) \equiv
\frac{1}{3 f^2} t_{\pi N}\rho
\end{equation}
for each pion line.
Next we separate the on-shell and off-shell parts of $t_{\pi N}$. For the
on-shell part we get
\begin{equation}
\label{swaveinsertion5}
\frac{1}{3 f^2} (\frac{4 c_1}{f^2}m_{\pi}^2-\frac{2 c_2}{f^2}\omega(q)^2-
\frac{2 c_3}{f^2}m_{\pi}^2) \rho \,\,\, ,
\end{equation}
which compared at threshold
to the free $t_{\pi\pi}$ amplitude, $t_{\pi\pi}= -\frac{1}{f^2} \frac{7}{2}
m_{\pi}^2$, gives 
\begin{equation}
\label{swaveinsertion6}
\frac{\delta t_{\pi\pi}}{t_{\pi\pi}} \simeq \frac{1}{21 f^2}
(8 c_1 - 4 c_2 - 4 c_3) \rho \,\,\, ,
\end{equation}
which with respect to Eq. (\ref{swaveinsertion3}) gets a reduction of a factor
21, plus an extra reduction from the near on-shell cancellation of the isoscalar
$t_{\pi N}$. Hence, this correction is negligible and we take advantage of this
large reduction factor $21$ to also neglect the part involving simultaneously
the off-shell parts of $t_{\pi\pi}$ and $t_{\pi N}$.

In order to proceed we have to decide upon the $c_i$ coefficients to be used. It
is well known that the Lagrangian of Eq. (\ref{LpiN2}) leads to a part of
$p-$wave pion selfenergy \cite{Kirchbach:1996xy}, but we are explicitly taking a
$p-$wave selfenergy insertion 
accounting for $ph$ and $\Delta h$ excitations. There is a work 
which uses the same Lagrangian of Eq. (\ref{LpiN2}), and
in addition takes into account explicitly the $\Delta$ degrees of freedom
\cite{Fettes:2000bb}. Thus, we stick to the values of the $c_i$ coefficients
obtained there from two fits, with and without using the $\sigma$ term as a
constraint, shown in Table \ref{ci}. For comparison, the values of the
coefficients $c_i$ without 
including the $\Delta$ are of the order of $c_1=-1.53$ GeV$^{-1}$, $c_2=3.22$
GeV$^{-1}$ and $c_3=-6.20$ GeV$^{-1}$ \cite{Fettes:1998ud}.
\begin{table}[ht]
\begin{center}
\begin{tabular}{|l|l|l|}
coef.& set I (GeV$^{-1}$) & set II (GeV$^{-1}$) \\ 
\hline
$c_1$ & $-0.35$ & $-0.32$ \\
\hline
$c_2$ & $-1.49$ & $-1.59$ \\
\hline
$c_3$ & $0.93$ & $1.15$ \\
\hline
\end{tabular}
\caption{\footnotesize{$c_i$ coefficients from Ref. \cite{Fettes:1998ud}.}}
\label{ci}
\end{center}
\end{table}
As stressed in \cite{Fettes:1998ud} the values of the coefficient $c_i$ with the
explicit contribution of the $\Delta$ are of natural order, while those obtained
without its consideration are too large and a source of problems in chiral
perturbative calculations \cite{Epelbaum:2003gr}. But in our case, as pointed
above, the choice is mandatory.

We can estimate the size of the correction of Eq. (\ref{correc_derivative})
at pion threshold, and taking advantage of the
reduction factor $1/3$ in the term $\frac{1}{3} \omega_i^2(q)$ in front of $s \simeq 4
m_{\pi}^2$, we approximate $\omega_i(q) \simeq m_{\pi}$. So we get
\begin{equation}
\label{delta}
\frac{\delta t_{\pi\pi}}{t_{\pi\pi}} = \frac{32}{21 f^2}(c_2 + c_3) \rho
+ \frac{10}{21 f^2}c_1 \rho \,\,\, ,
\end{equation}
which for the two sets of parameters of Table \ref{ci} gives
\begin{eqnarray}
\label{deltanum}
\frac{\delta t_{\pi\pi}}{t_{\pi\pi}}
&=& -0.154 \, \rho / \rho_0 \,\,\, \textrm{(set I)} \nonumber \\
&=& -0.124 \, \rho / \rho_0 \,\,\, \textrm{(set II)} \,\,\,.
\end{eqnarray}
Let us note that the correction is negative, reducing effectively the strength
of the $\pi\pi \to \pi\pi$ vertex in the medium. Note that should we have used
the values of $c_i$ without explicit $\Delta$ we would obtain a value for the
ratio of Eq. (\ref{delta}) of $-0.80 \rho / \rho_0$, certainly too large, but
also negative. 

Next we consider the contribution from Eq. (\ref{swaveinsertion3}). This
correction has opposite sign to the former one. When adding the two corrections
we find, again taking the threshold for comparison,
\begin{equation}
\label{deltatotal}
\frac{\delta t_{\pi\pi}}{t_{\pi\pi}} = -\frac{52}{21 f^2}(c_2 + c_3) \rho
+ \frac{10}{21 f^2} c_1 \rho \,\,\, ,
\end{equation}
which for the two sets of values of Table \ref{ci} gives
\begin{eqnarray}
\label{deltatotalnum}
\frac{\delta t_{\pi\pi}}{t_{\pi\pi}}
&=& 0.18 \, \rho / \rho_0 \,\,\, \textrm{(set I)} \nonumber \\
&=& 0.14 \, \rho / \rho_0 \,\,\, \textrm{(set II)} \,\,\,.
\end{eqnarray}
We can see that the sign of the correction is now reversed and, altogether, we
find now an effective increase of the $\pi\pi$ vertex in the medium by a
moderate amount.

Apart from the vertex corrections, we need to include the effect of the
on-shell $s-$wave pion selfenergy in  the pion propagators in the loops,
produced by the nucleon tadpole diagram. Since we have a broad range of pion
energies in the loop, we have used the $t \rho$ approximation for the $s-$wave
pion selfenergy and the amplitude $t$ has been  taken from the experimental fit
to data \cite{Arndt:1995bj}. This is a more realistic approach than to take the
expression from the model used here which gives a too large $s-$wave scattering
amplitude at high energies, and in any case produces a minor effect.

The considered corrections are included in the $\pi\pi$ amplitude by modifying
the kernel of the BS equation with the on-shell part of Eq.
(\ref{correc_derivative}) and  Eq. (\ref{swaveinsertion3}), namely
\begin{equation}
\label{BStad}
T = \frac{V+\delta t_{\pi\pi}^{(t)\,on}}
{1-(V+\delta t_{\pi\pi}^{(t)\,on}+\delta t_{\pi\pi}^{(st)})\,G}
\,\,\, ,
\end{equation}
and modifying the pion propagators in the calculation of the two-pion loop
function, $G$, as explained in Section \ref{sec:nucmed}.

\subsection{\label{sec:newmech}Vertex corrections from baryonic loops}

In the previous sections we have considered relevant medium effects, according
to the pion nucleus phenomenology, which describe correctly the pion in the
medium in a wide range of energies. These mechanisms lead to $s-$ and $p-$ wave
pion selfenergies in the propagators of the BS equation and some associated
vertex corrections. In Ref. \cite{Meissner:2001gz}, other vertex corrections to
the $\pi\pi$ amplitude which could provide some effect at low energies, where
the leading $p-$wave pion selfenergy is not so strong, were studied.

\begin{figure}
\begin{center}
\includegraphics[width=10cm,height=5cm]{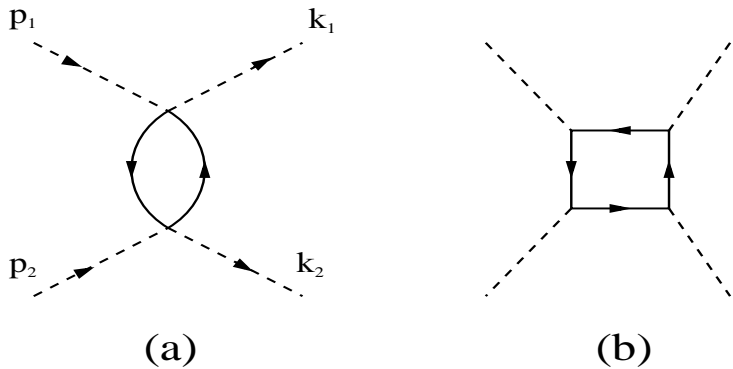}
\caption{\label{new_mech}(a) $ph$ bubble exchange in the $t$ channel; (b) Box
diagram.}
\end{center}
\end{figure}

The mechanisms considered in \cite{Meissner:2001gz} relevant for $\pi\pi$ $s-$wave
scattering, modifying the kernel of the Bethe
Salpeter equation in the $\pi \pi$ interaction, are shown in Fig. \ref{new_mech}.
The $\pi \pi$ isoscalar contribution in the $s$ channel for the $p h$ excitation
in the $t$ channel in Fig. \ref{new_mech}(a) is given, with the unitary normalization, by
\begin{equation}
-i t = -  \left( \frac{1}{4 f^2} \right) ^2 (p^0_1+k^{0}_1)(p^0_2+k^0_2) U(q) =
\tilde{t} \, U(q) 
\label{textra}
\end{equation}
where $U(q)$ is the ordinary Lindhard
function for $ph$ excitation, including a factor 2 of 
isospin (see Appendix of \cite{Oset:1990ey}). The second equation in Eq.
(\ref{textra}) defines $\tilde{t}$. 
We have neglected the isoscalar $\pi N$ amplitude 
in Eq. (\ref{textra}) since it is very small compared with the isovector one
\cite{Schroder:uq}.

The magnitude of $t$ was shown in \cite{Meissner:2001gz} to be comparable
to the $s-$wave $V$ of the lowest order chiral Lagrangian at densities of the 
order of the nuclear density. Yet, there are some observations to be made:
First, 
at pion threshold the diagram of Fig. \ref{new_mech}(a) is proportional to
$U(q^0=0,\vec{q}=\vec{0})$. This quantity is evaluated in \cite{Meissner:2001gz} using the
ordinary limit of the Lindhard function at $q^0=0$ and $|\vec{q}| \to 0$, which
is finite and larger in size than for any finite value of $|\vec{q}|$. This
limit is however quite different from the  value of the response
function at $\vec{q}=\vec{0}$ in finite nuclei which is strictly zero,
as already noted in \cite{Meissner:2001gz,Oset:xm,Oset:sm}.

We take into account the fact that the isovector $\pi N$ amplitude reflects
the exchange of a $\rho$ in the $t$ channel \cite{Ericson:gk} and multiply
$(4f^2)^{-2}$ by a factor reflecting the two $\rho$ propagators,
$F(q)=(M_{\rho}^2/(M_{\rho}^2+\vec{q}\,^2))^2$.

\begin{figure}
\begin{center}
\includegraphics[width=10cm,height=5cm]{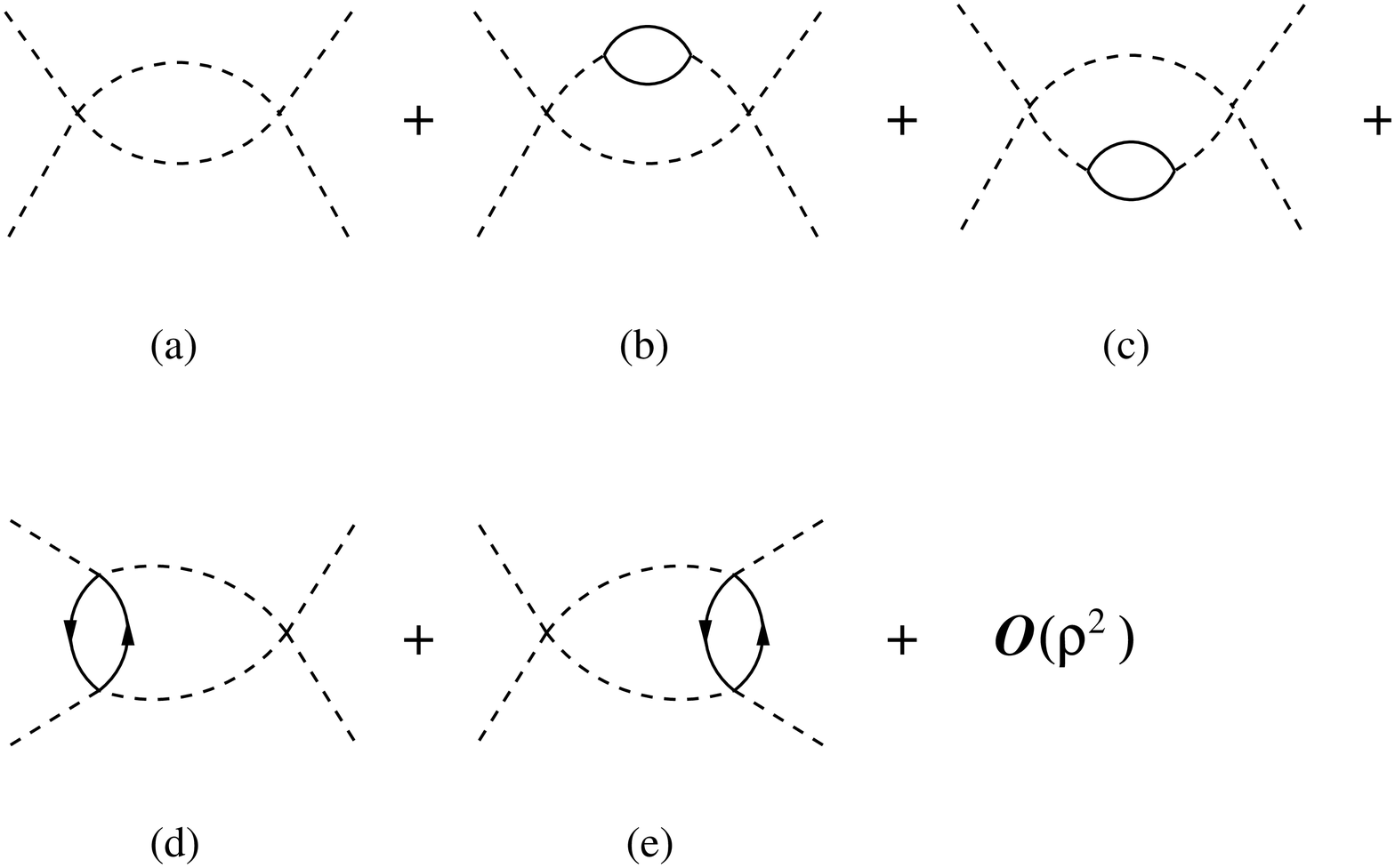}
\caption{\label{oneloopBS}Loop contributions to the Bethe Salpeter equation at first
order in the nuclear density, including the $t-$channel $ph$ excitation.}
\end{center}
\end{figure}

In order to estimate the importance of this contribution as compared to the
$p-$wave pion selfenergy insertions, we have evaluated the diagrams (a-e) in
Fig. \ref{oneloopBS}. Details of the calculation are given in Appendix II.
The results are shown in Fig. \ref{res:oneloopBS} for the
imaginary part of the resulting amplitude. We find that the contribution of
diagrams (d,e) is smaller than the changes produced by the insertion of the
$p-$wave pion selfenergy in the pion propagators. Similar results are found for
the real part of the amplitude.

\begin{figure}
\begin{center}
\includegraphics[width=0.7\textwidth]{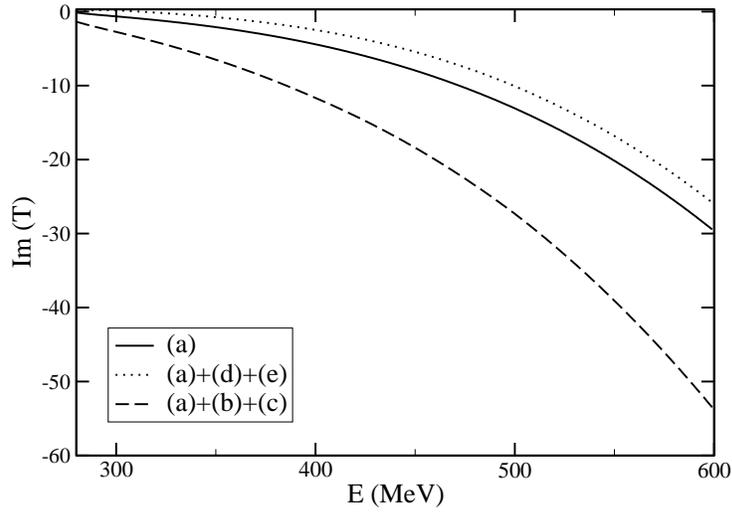}
\caption{\label{res:oneloopBS}Imaginary part of the $\pi\pi$ amplitude from the
terms in Fig. \ref{oneloopBS}, as indicated in the legend. The
calculation for the dashed and dotted lines is done for $\rho=\rho_0/2$.}
\end{center}
\end{figure}

The consideration of this mechanism in the BS equation proceeds by adding the
tree level term and modifying the kernel in the loop terms with an effective
potential $\delta V$ defined as
\begin{equation}
\frac{\delta V(s,\rho)}{V} = \frac{{\cal F}(s,\rho)}{V G V} \,\,\, ,
\label{Veff}
\end{equation}
where ${\cal F}(s,\rho)$ is the amplitude corresponding to diagram (b) and $VGV$
gives the amplitude of diagram (a) in Fig. \ref{oneloopBS}.
In this sense, by
substituting $V$ by $V+\delta V$ in the $\pi\pi$ vertex, the loop function of
diagram in Fig. \ref{oneloopBS}(a) would account correctly for all the diagrams
(a-e) at the first order in the nuclear density.

One of the reasons for the small size of this contribution is that the
Lindhard function behaves
roughly as $q^{-2}$ for large values of $q$ and we should expect a large
cancellation of this piece in the loops. This would be in contrast with the
$ph$ excitations leading to the $p-$wave $\pi$ selfenergy in Fig.
\ref{oneloopBS}(b,c), since
there one has the combination $\vec{q}\,^2 U(q)$ and a priori this type of $ph$
excitation should be more important, as it is indeed the case. Thus, the
$t-$channel $ph$ exchange mechanism leads to a sizeable correction to the tree
level $\pi\pi$ scattering amplitude and a small vertex correction in the
calculation of the loops appearing in the unitarization procedure\footnote{This
mechanism would play an even smaller role in the position of the $\sigma$ pole
\cite{VicenteVacas:2002se},
which is determined by the vanishing of the denominator of the BS solution,
where the tree level term does not appear.}.

Next we consider the box diagram of Fig. \ref{new_mech}(b). This term was found
to be smaller in strength than the $ph$ exchange in \cite{Meissner:2001gz},
particularly at small energies, where the $p-$wave character of the vertices
made the contribution negligible. 
The consideration of this mechanism at the pion loop level, necessary to include
it in the BS equation,
makes its contribution small since, apart from the reduction of the box diagram
for large values of $q$, there is a further cancellation of terms as we show in
Appendix III. Similar analytic treatments are done in
\cite{Herrmann:1993za,Cabrera:2000dx}. For all these reasons this contribution
should be even smaller than the one previously evaluated and one can safely
neglect it for practical purposes.

\begin{figure}
\begin{center}
\includegraphics[width=0.7\textwidth]{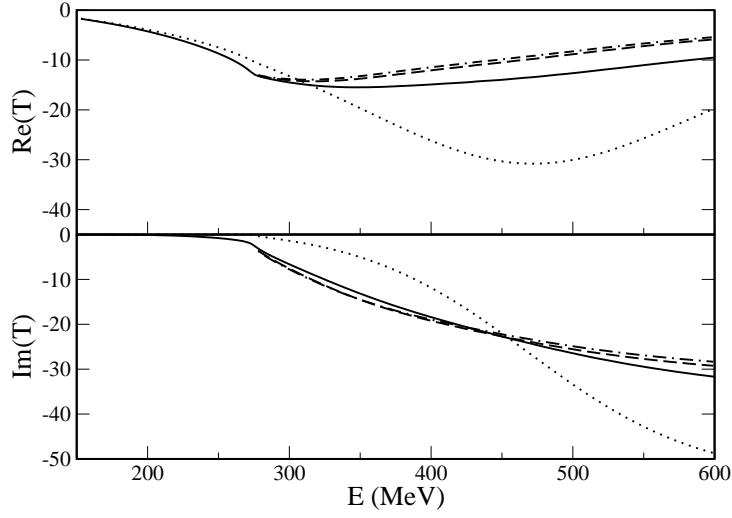}
\caption{\label{fig:tadeffect}Real and imaginary parts of $T$, as obtained in
Eq. (\ref{BStad}), for the two sets of parameters in Table \ref{ci} (set I,
dash-dotted line, set II, dashed line) and $\rho=\rho_0/2$. The solid line
corresponds to the result of the model in Sec. \ref{sec:nucmed} and the dotted
line is the result in vacuum.}
\end{center}
\end{figure}

\section{Results}

\begin{figure}
\begin{center}
\includegraphics[width=0.7\textwidth]{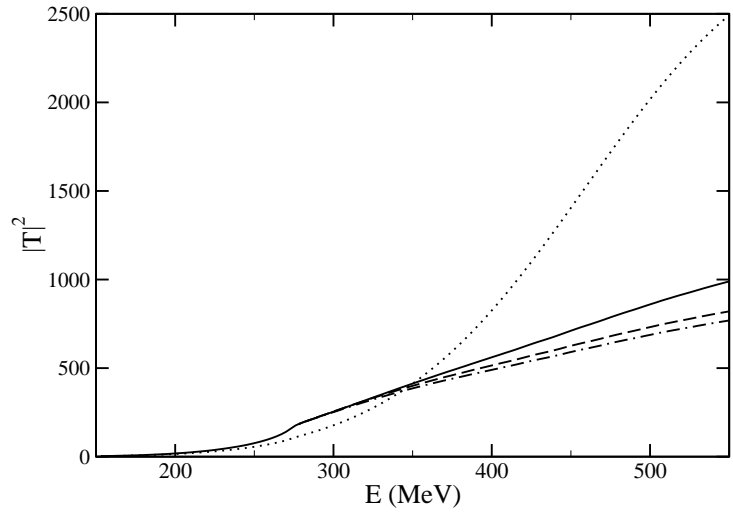}
\caption{\label{fig:tadeffectT2}Squared modulus of $T$. Lines as in Fig.
\ref{fig:tadeffect}.}
\end{center}
\end{figure}

We solve the BS equation including the corrections discussed in Sec.
\ref{sec:tadpoles}, as they appear in Eq. (\ref{BStad}). The results are shown in
Fig. \ref{fig:tadeffect}. The new terms considered modify little the
results from Ref. \cite{Chiang:1998di}. In comparison, the imaginary part of the
$\pi\pi$ amplitude exhibits a small increase of strength at low invariant
energies whereas the real part decreases over all the calculated range of
energies. Altogether, the basic effect of the nuclear medium, as found in Ref.
\cite{Chiang:1998di}, is a strong depletion of the interaction at energies
around $500$~MeV, where the vacuum $\sigma$ pole is found, and some accumulation
of strength close to the $2\pi$ threshold, as it can be seen in Fig.
\ref{fig:tadeffectT2}, where the squared modulus of the amplitude is 
depicted.

\begin{figure}
\begin{center}
\includegraphics[width=0.7\textwidth]{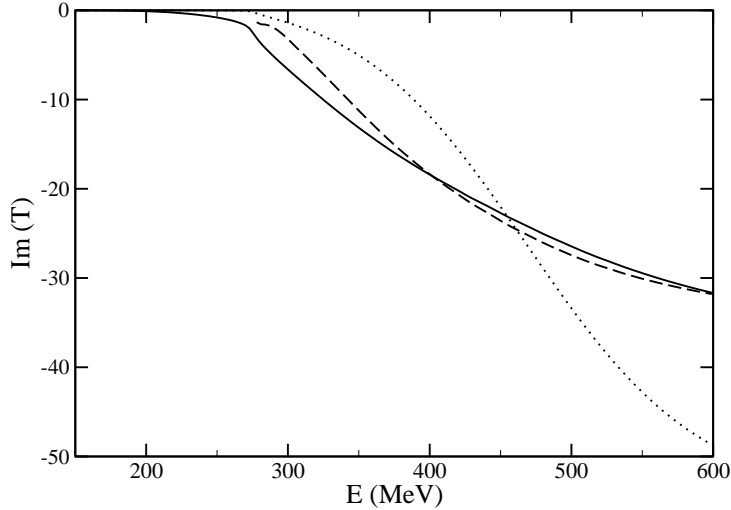}
\caption{\label{fig:newmech}Imaginary part of $T$ including the mechanism
described in Sec. \ref{sec:newmech} at $\rho=\rho_0/2$ (dashed line). The solid line
corresponds to the result of the model in Sec. \ref{sec:nucmed} and the dotted
line is the result in vacuum.}
\end{center}
\end{figure}

The contribution of the terms discussed in Sec. \ref{sec:newmech} is shown in
Fig. \ref{fig:newmech} for the imaginary part of the $\pi\pi$ amplitude. We find
a strong reduction of the amplitude at energies close to the $2\pi$ threshold,
basically produced by the repulsive tree level term in Fig. \ref{new_mech}(a). At
these energies the amplitude stays closer to the vacuum case. A similar
reduction of the nuclear medium effects as compared to the results of Ref.
\cite{Chiang:1998di} is found in the real part of the amplitude.

\begin{figure}
\begin{center}
\includegraphics[width=0.7\textwidth]{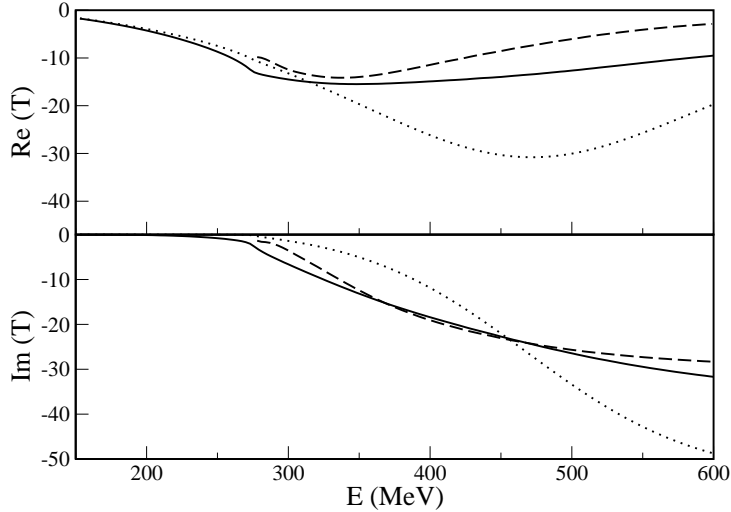}
\caption{\label{fig:all}Real and imaginary parts of $T$ including the
mechanisms described in Secs. \ref{sec:tadpoles} and \ref{sec:newmech} at
$\rho=\rho_0/2$, using set I (dashed line). The solid line corresponds to the result of the
model in Sec. \ref{sec:nucmed} and the dotted line is the result in vacuum.}
\end{center}
\end{figure}

Finally, we have included together the contributions of the tadpole terms,
Section \ref{sec:tadpoles}, and the $t-$channel $ph$ exchange, Section
\ref{sec:newmech}, in the BS equation, and the results are  depicted in Fig.
\ref{fig:all} for the real and imaginary parts of the $\pi\pi$ amplitude. We
observe, compared to the model of Ref. \cite{Chiang:1998di} in which the basic
medium effect is due to the $p-$wave pion selfenergy, a considerable reduction
of strength close to the two pion threshold. The global effect in both
calculations is still a sizable depletion of the interaction at higher energies
and a certain accumulation of strength below the $\sigma$ pole position in
vacuum which, as suggested in \cite{VicenteVacas:2002se}, could be reflecting a
change in the $\sigma$ pole position to lower energies as a function of the
nuclear density.

\section{Conclusions}

In summary, we have considered in this work the contribution of some new terms
to the $\pi\pi$ interaction in the scalar isoscalar channel at finite
densities, starting from a previous work \cite{Chiang:1998di,Oset:2000ev} in
which only medium effects associated to the $p-$wave pion selfenergy had been
accounted for.

Tadpole insertions, sometimes advocated as a possible source of a large
attraction, have been shown to affect little the $\pi\pi$ amplitude once the
Bethe Salpeter equation is solved. This is partly due to certain cancellations
which take place between vertices and internal pion propagator insertions.

  We have also taken into account new terms in the driving kernel of the Bethe
Salpeter equation, which have been found important in a study based on a
chiral power counting in the many body problem.   We could see that these new
terms, although large at tree level, when appearing inside loops were not as
important as one could guess from their comparison with the lowest order chiral
$\pi \pi$ amplitude in the case that all pions are on shell.  As a consequence,
their consideration barely changed the results for the $\pi\pi$ interaction in
the medium.

Altogether, the final results are quite similar to those obtained previously in
\cite{Chiang:1998di,Oset:2000ev}, namely a strong reduction of the interaction
at energies around 400~MeV and beyond, and some increase of strength around the
2$\pi$ threshold.
This confirms the leading role of the strong $p-$wave pion selfenergy in the
medium modification of the $\pi\pi$ interaction in the scalar isoscalar channel.
These results are also satisfactory because a prediction on the $(\gamma,2\pi)$
reaction in nuclei \cite{Roca:2002vd} based on the previous calculation
\cite{Chiang:1998di,Oset:2000ev} of the two-pion final state interaction has
been later confirmed by experimental data \cite{Messchendorp:2002au}. The much
larger medium effects obtained at threshold energies in other approaches are
incompatible with the observed effect in the $(\gamma,2\pi)$ reaction.

\section*{Acknowledgements}
This work is partly supported by DGICYT contract no. 
BFM2003-00856. D.~Cabrera acknowledges financial support from MEC.

\section*{Appendix I}
We quote in this section the Lindhard function, with an energy gap $\Delta$,
separated into the direct and crossed contributions, $U=U_d+U_c$. From
\cite{Oset:sm} we have 
\begin{equation}
U_d(q^0,\vec{q},\Delta;\rho) = 4 \int \frac{d^3 p}{(2\pi)^3} 
\frac{n(\vec{p}) \lbrack 1 - n(\vec{p}+\vec{q}) \rbrack}
{q^0 + \varepsilon (\vec{p}) - \varepsilon (\vec{p}+\vec{q}) - \Delta + i \epsilon}
\label{U_dgap}
\end{equation}
and $U_c(q^0,\vec{q},\Delta;\rho) \equiv U_d(-q^0,\vec{q},\Delta;\rho)$. In the
following we shall use the definitions
\begin{eqnarray}
x = \frac{q}{k_{F}} \,\,\, , \,\,\, \nu = \frac{2Mq^0}{k_F^2} \nonumber \\
\delta = \frac{2M\Delta}{k_F^2} \,\,\, , \,\,\, \rho=\frac{2}{3\pi^2}k_F^3
\label{defs}
\end{eqnarray}
with $M$ the mass of the nucleon, $k_F$ the Fermi momentum and $q \equiv
|\vec{q}|$. Once the integration in Eq. (\ref{U_dgap}) is done, the real
part of $U_d$ reads, for $x \leq 2$,
\begin{eqnarray}
\textrm{Re} \, U_d(q^0,\vec{q},\Delta;\rho) = - \frac{2Mk_F}{\pi^2} \frac{1}{2x}
\bigg \lbrace \frac{x}{2} - \frac{\nu - \delta}{4} 
+ \frac{\nu - \delta}{2} \ln \bigg | \frac{\nu-\delta+x^2-2x}{\nu-\delta}\bigg |
\nonumber \\
+ \frac{1}{2}\left[ 1-\frac{1}{4} \left( \frac{\nu-\delta}{x}-x \right) ^2
\right] \ln \bigg | \frac{\nu-\delta-x^2-2x}{\nu-\delta+x^2-2x} \bigg | \bigg
\rbrace
\label{Reless2}
\end{eqnarray}
and, for $x > 2$,
\begin{eqnarray}
\textrm{Re} \, U_d(q^0,\vec{q},\Delta;\rho) = - \frac{2Mk_F}{\pi^2}\frac{1}{2x}
\bigg \lbrace \frac{-\nu+\delta+x^2}{2x} \nonumber \\
+ \frac{1}{2}\left[ 1 - \frac{1}{4} \left( \frac{\nu-\delta}{x}-x\right)^2
\right] \ln
\bigg | \frac{\nu-\delta-x^2-2x}{\nu-\delta-x^2+2x} \bigg | \bigg \rbrace 
\,\,\, .
\label{Regt2}
\end{eqnarray}
The imaginary part of $U_d$ is given by
Im $U_d(q^0,\vec{q},\Delta;\rho)=$ Im $\tilde{U}(q^0-\Delta,\vec{q};\rho)
\Theta(q^0-\Delta)$, with
\begin{eqnarray}
\textrm{Im}\, \tilde{U}(q^0,\vec{q};\rho)= -\frac{3}{4} \pi \rho \frac{M}{q k_F}
\lbrack (1-z^2)\Theta(1-|z|) - (1-z'\,^2)\Theta(1-|z'|) \rbrack \frac{q^0}{|q^0|}
\,\,\, ,
\label{Utilda}
\end{eqnarray}
where $\Theta$ is the Heaviside step function and the $z$, $z'$ variables are
defined as 
\begin{equation}
z=\frac{M}{q k_F}\left[ q^0-\frac{q^2}{2M}\right] \,\,\, , \,\,\,
z'=\frac{M}{q k_F}\left[ -q^0-\frac{q^2}{2M}\right] \,\,\, .
\end{equation}

\section*{Appendix II}
\begin{figure}
\begin{center}
\includegraphics[width=6cm]{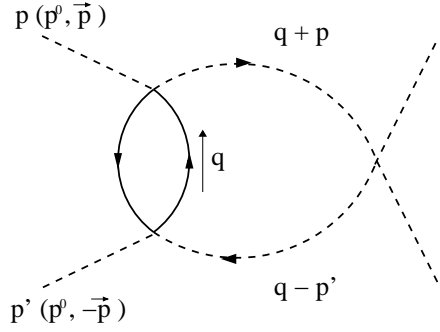}
\caption{\label{loop2}Loop contribution of the $ph$ exchange in the $t$
channel.}
\end{center}
\end{figure}
The amplitude corresponding to the diagram in Fig. \ref{loop2} is given by
\begin{equation}
-i T = \int \frac{d^4q}{(2\pi)^4} (-i \tilde{t}) \frac{i}{(q+p)^2 - m_{\pi}^2 +
i\epsilon} \frac{i}{(q-p')^2 - m_{\pi}^2 + i\epsilon} (-i V(s)) \, i U(q) \,\,\,
.
\label{new_T}
\end{equation}
In order to perform the integral it is most useful to separate $U(q)$ into the
direct and crossed parts, $U(q)=U_d(q)+U_c(q)$, given their different analytical
structure.
\begin{figure}
\begin{center}
\includegraphics[width=13cm]{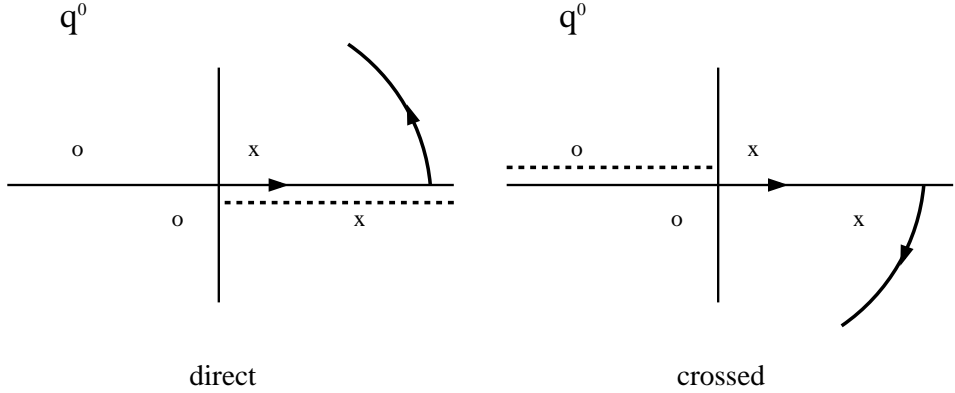}
\caption{\label{analyt}Analytical structure of the integrand in Eq.
(\ref{new_T}). The poles of the pion propagators are represented by '$x$' and
'$o$' symbols, and the dotted lines correspond to the analytical cuts of the
Lindhard function. The arrows indicate the circuit used for the integration of
each term.} 
\end{center}
\end{figure}

In Fig. \ref{analyt} we depict the pole and cut structure for the different
terms and the path followed for the integration in the complex plane. The poles
are located at
\begin{eqnarray}
q^0 = -p^0 + \omega(\vec{p}+\vec{q}) - i\epsilon \,\,\, , \,\,\,
q^0 = -p^0 - \omega(\vec{p}+\vec{q}) + i\epsilon \nonumber \\
q^0 = p^0 + \omega(\vec{p}+\vec{q}) - i\epsilon \,\,\, , \,\,\,
q^0 = p^0 - \omega(\vec{p}+\vec{q}) + i\epsilon \,\,\, .
\label{poles}
\end{eqnarray}
The integration over the $q^0$ variable is done by closing the contour in the
complex plane in the upper half plane for the $U_d$ part and in the lower half
plane for the $U_c$ part.

The result of the integration is
\begin{eqnarray}
T = - \left( \frac{1}{4f^2}\right)^2 (2p^0)^2 V(s)
\int \frac{d^3q}{(2\pi)^3} \frac{1}{4\omega^2} \bigg \lbrace
\frac{U_c(p^0+\omega,\vec{q})}{p^0+\omega} \nonumber \\
-\frac{U_d(p^0-\omega,\vec{q})}{p^0-\omega+i\epsilon}
+ \frac{U_d(p^0-\omega,\vec{q})-U_c(p^0+\omega,\vec{q})}{p^0} \bigg \rbrace \bigg (
\frac{M_{\rho}^2}{M_{\rho}^2+\vec{q}\,^2} \bigg ) ^2 \,\,\, ,
\end{eqnarray}
where $\omega \equiv \omega(\vec{p}+\vec{q})$ and we have explicitly written the
$\rho$ meson exchange form factor arising from each $\pi\pi NN$ vertex.

Let us note that we have factorized the $\pi N \to \pi N$ vertex on shell. This
is done in analogy to what is done in \cite{Oset:1997it} where one shows that the
off shell part can be cast into a renormalization of the lowest order diagram
(no meson loop in this case). An alternative justification using dispersion
relations, which require only the on shell information, is given in
\cite{Oller:2000fj}.

\section*{Appendix III}
\begin{figure}
\begin{center}
\includegraphics[width=5cm]{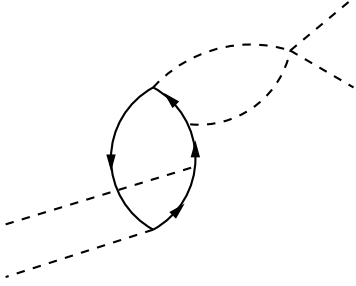}
\caption{\label{box}Box diagram with two of the pions as a part of a loop.}
\end{center}
\end{figure}

\begin{figure}
\begin{center}
\includegraphics[width=7cm]{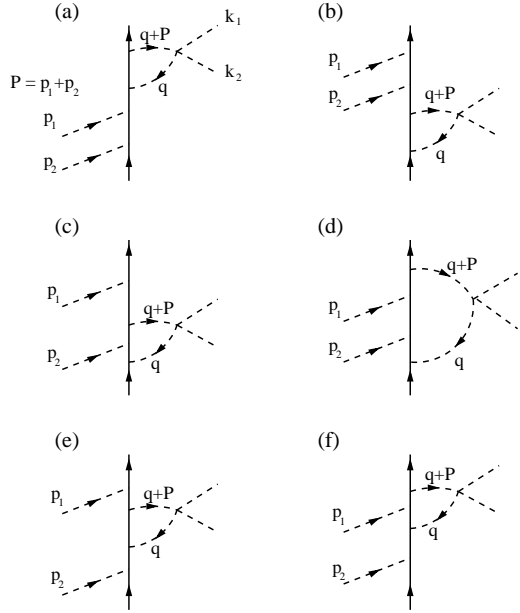}
\caption{\label{box_orderings}Set of different time orderings of diagram in Fig.
\ref{box}. The initial and final nucleon lines correspond to a hole
propagator.}
\end{center}
\end{figure}

We evaluate the loop function of Fig. \ref{box} containing the box
diagram of Fig. \ref{new_mech}(b)
 plus all the different time orderings,
which we can see in Fig. \ref{box_orderings}. In all the diagrams the internal nucleon
lines are particle lines. This means we are taking only the terms 
of order $\rho$, which are obtained when the external lines
are folded to give a single hole line in Fig. \ref{box_orderings}.

{\bf Diagrams (a), (b)}. For diagrams (a), (b) in Fig. \ref{box_orderings} for
$\vec{P}=\vec{p}_1+\vec{p}_2=\vec{0}$ the intermediate nucleon line after the
two pion vertices has the same momentum as the hole line (belonging to the
Fermi sea) and hence they both vanish. 

Next we observe some strong
cancellations in the other diagrams. The set of the two meson propagators, which
is common to all of them, can be
written as
\begin{equation}
\frac{1}{2 P^0 \omega} \bigg \lbrace
\frac{1}{q^0-\omega+i\epsilon} \; \frac{1}{q^0+P^0+\omega-i\epsilon}
- \frac{1}{q^0+\omega-i\epsilon} \; \frac{1}{q^0+P^0-\omega+i\epsilon}
\bigg \rbrace \,\,\, ,
\label{props}
\end{equation}
with $\omega=\omega(\vec{q})$.

{\bf Diagram (c)}. The diagram (c) contains three nucleon
propagators. By making a heavy baryon approximation and neglecting the kinetic
energy of the nucleons we find for the product
\begin{equation}
\frac{1}{-p_1^0} \; \frac{1}{q^0+p_2^0+i\epsilon} \; \frac{1}{q^0+i\epsilon} \,\,\, .
\label{approx_c}
\end{equation}
Hence, multiplying this by the pion propagators and closing the contour on the
upper half plane to perform the $q^0$ integration, we find that the integral is
\begin{equation}
A \frac{1}{p_1^0} \bigg \lbrace -\frac{1}{p_1^0+\omega}\; \frac{1}{P^0+\omega}
+ \frac{1}{\omega} \; \frac{1}{p_2^0-\omega+i\epsilon} \bigg \rbrace \,\,\ .
\label{integ_c}
\end{equation}
There, the first term, which comes from the negative energy components of the
mesons, is small and has no imaginary part. The second term can lead to an
imaginary part and a more sizeable real part from the principal value.

{\bf Diagram (e)}. The set of nucleon propagators in the heavy baryon
approximation is now
\begin{equation}
\frac{1}{p_2^0} \; \frac{1}{p_2^0+q^0+i\epsilon} \; \frac{1}{-p_1^0}
\label{approx_e}
\end{equation}
and hence by closing the contour in the upper half of the complex $q^0$
plane we find for the $q^0$ integration
\begin{equation}
A \frac{1}{p_1^0}\bigg \lbrace \frac{1}{p_2^0} \; \frac{1}{p_1^0+\omega}
- \frac{1}{p_2^0} \; \frac{1}{p_2^0-\omega+i\epsilon} \bigg \rbrace \,\,\, ,
\label{integ_e}
\end{equation}
with the same $A$ as in Eq. (\ref{integ_c}).
The first term is again small, coming from the negative energy components of the
pions, and has opposite sign to the first term from diagram (c). The second term
above is the same but with opposite sign to the second term of diagram (c) at
$\omega=p_2^0$, which is the singular point. Hence there are strong
cancellations in the principal part of the integral and the imaginary part from
this source vanishes.

{\bf Diagram (d)}. Repeating the same arguments as above we find now
\begin{equation}
-A \frac{1}{\omega}\bigg \lbrace \frac{1}{p_1^0+\omega} \; \frac{1}{P^0+\omega}
+ \frac{1}{P^0-\omega+i\epsilon} \; \frac{1}{p_2^0-\omega+i\epsilon} \bigg \rbrace \,\,\, .
\label{integ_d}
\end{equation}

{\bf Diagram (f)}. For this diagram we find
\begin{equation}
A \frac{1}{p_1^0}\bigg \lbrace \frac{1}{\omega} \; \frac{1}{p_2^0+\omega}
- \frac{1}{P^0-\omega+i\epsilon} \; \frac{1}{p_1^0-\omega+i\epsilon} \bigg \rbrace \,\,\, .
\label{integ_f}
\end{equation}

Once again the first two terms from (d), (f), coming from the negative energy
part of the pion propagators, give a small contribution and partly cancel, and
the second terms which provide an imaginary part and a larger real part from the
principal value, also show cancellations. Indeed for $p_1^0=p_2^0=\omega$ the
imaginary parts corresponding to the poles $p_1^0=p_2^0=\omega$ cancel and the
real parts from the principal value would also largely cancel. At the
$P^0=\omega$ pole the cancellation would only be partial.

We thus see that when considering all the time orderings for the coupling of the
two pions and the loop with the two pion propagators there are large
cancellations of terms. In addition we have the $(p_i / M)^2$ factor of the
$p-$wave couplings for the initial pions, which make this contribution small at
small momenta of the pions. We have looked at strong cancellations of terms in
the heavy baryon approximation, which holds for small values of momenta. At
large momenta 
we must note that we have two extra nucleon propagators which bring two
extra powers of $q$ in the denominator, with respect to the
ordinary Lindhard function. This makes up for the two extra $p-$wave vertices, and
hence we have a similar behaviour altogether as the one of Fig. \ref{loop2}
which lead to small contributions when evaluated into the loop. All these
elements discussed above would render this piece far smaller than the ones
of Fig. \ref{oneloopBS}(d,e) and, given the smallness of the effects found there, this
can also be neglected.


\begin{thebibliography}{99}



\bibitem{Hatsuda:1999kd}
T.~Hatsuda, T.~Kunihiro and H.~Shimizu,
Phys.\ Rev.\ Lett.\  {\bf 82} (1999) 2840.

\bibitem{Jido:2000bw}
D.~Jido, T.~Hatsuda and T.~Kunihiro,
Phys.\ Rev.\ D {\bf 63} (2001) 011901
[arXiv:hep-ph/0008076].


\bibitem{Schuck:1988jn}
P.~Schuck, W.~Norenberg and G.~Chanfray,
Z.\ Phys.\ A {\bf 330} (1988) 119 .

\bibitem{Rapp:1996ir}
R.~Rapp, J.~W.~Durso and J.~Wambach,
Nucl.\ Phys.\ A {\bf 596} (1996) 436 
[arXiv:nucl-th/9508026].

\bibitem{Aouissat:1995sx}
Z.~Aouissat, R.~Rapp, G.~Chanfray, P.~Schuck and J.~Wambach,
Nucl.\ Phys.\ A {\bf 581} (1995) 471 
[arXiv:nucl-th/9406010].

\bibitem{Chiang:1998di}
H.~C.~Chiang, E.~Oset and M.~J.~Vicente-Vacas,
Nucl.\ Phys.\ A {\bf 644} (1998) 77 
[arXiv:nucl-th/9712047].

\bibitem{Aouissat:2000ss}
Z.~Aouissat, G.~Chanfray, P.~Schuck and J.~Wambach,
Phys.\ Rev.\ C {\bf 61} (2000) 012202.

\bibitem{Davesne:2000qj}
D.~Davesne, Y.~J.~Zhang and G.~Chanfray,
Phys.\ Rev.\ C {\bf 62} (2000) 024604
[arXiv:nucl-th/9909032].

\bibitem{Bonutti:1996ij}
F.~Bonutti {\it et al.}  [CHAOS Collaboration],
Phys.\ Rev.\ Lett.\  {\bf 77} (1996) 603.

\bibitem{Bonutti:1998zw}
F.~Bonutti {\it et al.}  [CHAOS Collaboration],
Nucl.\ Phys.\ A {\bf 638} (1998) 729. 

\bibitem{Camerini:1993ac}
P.~Camerini, N.~Grion, R.~Rui and D.~Vetterli,
Nucl.\ Phys.\ A {\bf 552} (1993) 451 
[Erratum-ibid.\ A {\bf 572}  (1993) 791].

\bibitem{bonutti}
F.~Bonutti {\it et al.}  [CHAOS Collaboration],
Phys.\ Rev.\  {\bf C60}  (1999) 018201.

\bibitem{Starostin:2000cb}
A.~Starostin {\it et al.}  [Crystal Ball Collaboration],
Phys.\ Rev.\ Lett.\  {\bf 85} (2000) 5539.

\bibitem{Messchendorp:2002au}
J.~G.~Messchendorp {\it et al.},
Phys.\ Rev.\ Lett.\  {\bf 89} (2002) 222302
[arXiv:nucl-ex/0205009].

\bibitem{Roca:2002vd}
L.~Roca, E.~Oset and M.~J.~Vicente Vacas,
Phys.\ Lett.\ B {\bf 541} (2002) 77
[arXiv:nucl-th/0201054].

\bibitem{Muhlich:2004zj}
P.~Muhlich, L.~Alvarez-Ruso, O.~Buss and U.~Mosel,
Phys.\ Lett.\ B {\bf 595} (2004) 216
[arXiv:nucl-th/0401042].

\bibitem{Dobado:1990qm}
A.~Dobado, M.~J.~Herrero and T.~N.~Truong,
Phys.\ Lett.\ B {\bf 235} (1990) 134.

\bibitem{Dobado:1993ha}
A.~Dobado and J.~R.~Pelaez,
Phys.\ Rev.\ D {\bf 47} (1993) 4883 
[arXiv:hep-ph/9301276].

\bibitem{Oller:1998ng}
J.~A.~Oller, E.~Oset and J.~R.~Pelaez,
Phys.\ Rev.\ Lett.\  {\bf 80} (1998) 3452 
[arXiv:hep-ph/9803242].

\bibitem{Oller:1999hw}
J.~A.~Oller, E.~Oset and J.~R.~Pelaez,
Phys.\ Rev.\ D {\bf 59} (1999) 074001 
[Erratum-ibid.\ D {\bf 60} (1999) 099906]
[arXiv:hep-ph/9804209].

\bibitem{Oller:1999zr}
J.~A.~Oller and E.~Oset,
Phys.\ Rev.\ D {\bf 60} (1999) 074023 
[arXiv:hep-ph/9809337].

\bibitem{Oller:1997ti}
J.~A.~Oller and E.~Oset,
Nucl.\ Phys.\ A {\bf 620} (1997) 438 
[Erratum-ibid.\ A {\bf 652} (1997) 407]
[arXiv:hep-ph/9702314].

\bibitem{Gasser:1985ux}
J.~Gasser and H.~Leutwyler,
Nucl.\ Phys.\ B {\bf 250} (1985) 517.

\bibitem{Oset:2000ev}
E.~Oset and M.~J.~Vicente Vacas,
Nucl.\ Phys.\ A {\bf 678} (2000) 424
[arXiv:nucl-th/0004030].

\bibitem{Nieves:2000bx}
J.~Nieves and E.~Ruiz Arriola,
Nucl.\ Phys.\ A {\bf 679} (2000) 57
[arXiv:hep-ph/9907469].

\bibitem{Nieves:1999hp}
J.~Nieves and E.~Ruiz Arriola,
Phys.\ Lett.\ B {\bf 455} (1999) 30
[arXiv:nucl-th/9807035].

\bibitem{Oset:1990ey}
E.~Oset, P.~Fernandez de Cordoba, L.~L.~Salcedo and R.~Brockmann,
Phys.\ Rept.\  {\bf 188} (1990) 79.

\bibitem{Chanfray:1999nn}
G.~Chanfray and D.~Davesne,
Nucl.\ Phys.\ A {\bf 646} (1999) 125.

\bibitem{Bernard:1995dp}
V.~Bernard, N.~Kaiser and U.~G.~Meissner,
Int.\ J.\ Mod.\ Phys.\ E {\bf 4} (1995) 193 
[arXiv:hep-ph/9501384].

\bibitem{Meissner:2001gz}
U.~G.~Meissner, J.~A.~Oller and A.~Wirzba,
Annals Phys.\  {\bf 297} (2002) 27
[arXiv:nucl-th/0109026].

\bibitem{Schroder:uq}
H.~C.~Schroder {\it et al.},
Phys.\ Lett.\ B {\bf 469} (1999) 25.

\bibitem{Oset:xm}
E.~Oset and D.~Strottman,
Phys.\ Rev.\ Lett.\  {\bf 70} (1993) 146.

\bibitem{Oset:sm}
E.~Oset, D.~Strottman, H.~Toki and J.~Navarro,
Phys.\ Rev.\ C {\bf 48} (1993) 2395.

\bibitem{Ericson:gk}
T.~E.~Ericson and W.~Weise,
{\it  OXFORD, UK: CLARENDON (1988) 479 P. (THE INTERNATIONAL SERIES OF MONOGRAPHS ON PHYSICS, 74)}.

\bibitem{Herrmann:1993za}
M.~Herrmann, B.~L.~Friman and W.~Norenberg,
Nucl.\ Phys.\ A {\bf 560} (1993) 411.

\bibitem{Cabrera:2000dx}
D.~Cabrera, E.~Oset and M.~J.~Vicente Vacas,
Nucl.\ Phys.\ A {\bf 705} (2002) 90
[arXiv:nucl-th/0011037].

\bibitem{Arndt:1995bj}
R.~A.~Arndt, I.~I.~Strakovsky, R.~L.~Workman and M.~M.~Pavan,
Phys.\ Rev.\ C {\bf 52} (1995) 2120
[arXiv:nucl-th/9505040].

\bibitem{Oset:1997it}
E.~Oset and A.~Ramos,
Nucl.\ Phys.\ A {\bf 635} (1998) 99
[arXiv:nucl-th/9711022].

\bibitem{Oller:2000fj}
J.~A.~Oller and U.~G.~Meissner,
Phys.\ Lett.\ B {\bf 500} (2001) 263
[arXiv:hep-ph/0011146].

\bibitem{Thorsson:1995rj}
V.~Thorsson and A.~Wirzba,
Nucl.\ Phys.\ A {\bf 589} (1995) 633
[arXiv:nucl-th/9502003].

\bibitem{Oller:1998zr}
J.~A.~Oller and E.~Oset,
Phys.\ Rev.\ D {\bf 60} (1999) 074023
[arXiv:hep-ph/9809337].

\bibitem{Chanfray:nn}
G.~Chanfray and D.~Davesne,
Nucl.\ Phys.\ A {\bf 646} (1999) 125.

\bibitem{Kirchbach:1996xy}
M.~Kirchbach and A.~Wirzba,
Nucl.\ Phys.\ A {\bf 604} (1996) 395
[arXiv:nucl-th/9603017].

\bibitem{Fettes:2000bb}
N.~Fettes and U.~G.~Meissner,
Nucl.\ Phys.\ A {\bf 679} (2001) 629
[arXiv:hep-ph/0006299].

\bibitem{Fettes:1998ud}
N.~Fettes, U.~G.~Meissner and S.~Steininger,
Nucl.\ Phys.\ A {\bf 640} (1998) 199
[arXiv:hep-ph/9803266].

\bibitem{Epelbaum:2003gr}
E.~Epelbaum, W.~Gloeckle and U.~G.~Meissner,
Eur.\ Phys.\ J.\ A {\bf 19} (2004) 125
[arXiv:nucl-th/0304037].

\bibitem{VicenteVacas:2002se}
M.~J.~Vicente Vacas and E.~Oset,
arXiv:nucl-th/0204055.

\end{thebibliography}
\end{document}